\documentclass[aps,prd,reprint,nofootinbib]{revtex4-1}
\usepackage{blindtext}
\usepackage{mathtools}
\usepackage{xcolor}

\begin{document}
\title{Axion star nucleation in dark minihalos around primordial black holes}
\author{Mark P. Hertzberg$^{1*}$, Enrico D. Schiappacasse$^{2,3,4\dagger}$, Tsutomu T. Yanagida$^{4\ddagger}$}
\affiliation{$^1$Institute of Cosmology, Department of Physics and Astronomy, Tufts University, Medford, Massachusetts 02155, USA\\
$^2$ Department of Physics, P.O.Box 35 (YFL), FIN-40014 University of Jyv$\ddot{a}$skyl$\ddot{a}$, Finland\\
$^3$Helsinki Institute of Physics, P.O. Box 64, FIN-00014 University of Helsinki, Finland\\ 
$^4$Tsung-Dao Lee Institute and School of Physics and Astronomy, Shanghai Jiao Tong University, 
 200240 Shanghai, China }

\begin{abstract}
We consider a general class of axion models, including the QCD and string axion, in which the PQ symmetry is broken before or during inflation. Assuming the axion is the dominant component of the dark matter, we discuss axion star formation in virialized dark minihalos around primordial black holes through gravitational Bose-Einstein condensation. We determine the conditions for minihalos to kinetically produce axion stars before galaxy formation. Today, we expect up to $\sim 10^{17}$ ($\sim 10^9$) axion stars in a radius of 100 parsecs around the Sun for the case of the QCD (string) axion.
 \\ \\  
\end{abstract}

\maketitle

\section{Introduction}
By considering shortcomings  in the Standard Model
		of particle physics, the axion is one of the strongest dark matter candidates~\cite{Peccei:1977hh, Weinberg:1977ma,  Wilczek:1977pj, Kiritsis:2014yqa, PRESKILL1983127, Abbott:1982af, DINE1983137}.  In the scenario at which the PQ symmetry is broken after inflation, the axion field initially takes random values from one Hubble patch to the next leading to large isocurvature perturbations in the axion energy density at the QCD phase transition.  These overdensities will decouple from the Hubble flow and form the so-called axion miniclusters~\cite{Hogan:1988mp, Kolb:1993zz, Enander:2017ogx}. 

However, the scenario at which the PQ symmetry is broken after inflation is in tension with numerical simulations reported in~\cite{Kawasaki:2014sqa}.  The additional axion abundance coming from the decay of topological defects, 
significatively changes the usual axion abundance coming from the misalignment mechanism~\cite{DAVIS1986225, Lyth:1991bb}.   When the domain wall number ($N_{\text{DW}}$) is equal to the unity, the axion can be the cold dark matter in the Universe in the narrow mass range $m_{a} =  (0.8\times 10^{-4} - 1.4\times 10^{-2})\,\text{eV}$. If $N_{\text{DW}} > 1$, 
the QCD axion is excluded in the standard scenario. However, it may be rescued by including a \textit{bias} term in the potential of the PQ field
~\cite{Kawasaki:2014sqa}.  

In this paper, we consider a more general class of the axion model, including the string axion, where the PQ symmetry is broken during or before the inflation. In this case, we have the so-called isocurvature perturbation problem~\cite{Axenides:1983hj, Seckel:1985tj, Linde:1985yf, Linde:1990yj, Turner:1990uz, Lyth:1991ub}, but there are many solutions to this~(see, for example, \cite{Kawasaki:2013iha}). If this is the case, density fluctuations of the axion dark matter are sufficiently small.  

In particular, we discuss the formation of axion stars\footnote{An axion star is a kind of boson star (see~\cite{Liebling2012} for a review about boson stars and \cite{Horvat:2012aq, Choi:2019mva, Guerra:2019srj} for novel extensions) corresponding to self-gravitating bound states of an axion Bose-Einstein condensation~\cite{Guth:2014hsa, Schiappacasse:2017ham, Hertzberg:2018lmt, Visinelli:2017ooc}.} in dark minihalos around primordial black holes (PBHs) through gravitational Bose-Einstein condensation (BEC) in the kinetic regime.  PBHs~\cite{Hawking:1971ei, Carr:1974nx, Carr:1975qj, Kawasaki:1997ju, GarciaBellido:1996qt, Khlopov:2008qy} behave as cold dark
matter. Nowadays, their possible existence has been strongly revitalized since the first detection of two merging black holes by the LIGO-Virgo Collaboration~\cite{Abbott:2016blz}. Since PBHs are local overdensities in the dark matter distribution, they naturally act as seeds for dark matter structures formation. In the scenario at which the axion is the dominant component of the dark matter, dark minihalos will unavoidably grow around PBHs. If PBHs exist, this scenario is realized whatever the fundamental nature of original axion dark matter distribution is.

The kinetic formation of axion stars in these dark minihalos, where the axion field coherence length is much smaller than the halo radius, depends on the halo energy density as well as  the axion mass and velocity. Neglecting a weak logarithm dependence, the time scale for axion star nucleation runs as   $\tau_{\text{gr}}  \sim m_a^3 v_a^6 \rho_{\text{halo}}^{-2}$, where  $m_a$ and $v_a$ refer to the axion mass and velocity, respectively, and $\rho_{\text{halo}}$ is the halo energy density~\cite{Levkov:2018kau}.  In this paper, we show accretion of axion dark matter around PBHs is effective enough to achieve axion stars formation before dressed PBHs begin to interact with nonlinear structures.

\section{Dark Minihalos around PBHs}
 Primordial black holes which are formed with a mass $M_{\text{PBH}} \gtrsim 10^{15}$ gram do not evaporate but begin to  form compact dark matter halos by accreting the surrounding axion dark matter. 

Any overdensity within a sphere in an expanding Universe will seed the growth of a minihalo according to the theory of spherical gravitational collapse~\cite{Bertschinger:1985pd}. 
Under the assumption that each PBH is stationary and isolated, and dark matter background is initially in the Hubble flow, analytical and numerical calculations~\cite{Mack:2006gz} show PBH dark minihalos mainly grow during the matter-dominated era 
reaching up to $\sim10^2 M_{\text{PBH}}$  (in units of the central PBH mass, $M_{\text{PBH}})$. 
The dark minihalo mass and radius grow in terms of $M_{\text{PBH}}$ and redshift $z$ as~\cite{Mack:2006gz, Ricotti:2007au}
\begin{align}
M_{\text{halo}} (z) &= 3\left(\frac{1000}{1+z}\right) M_{\text{PBH}}\,,\label{mh}\\
R_{\text{halo}} (z) = 0.019\, &\text{pc} \left( \frac{M_{\text{halo}}}{M_{\odot}} \right)^{1/3} \left( \frac{1000}{1+z} \right)\,.\label{rh}
\end{align}
Both expressions agree very well with calculations of the virial mass and radius performed in~\cite{Berezinsky:2013fxa}. 
These relations hold until the time of first galaxies formation at $z \sim 30$, when dressed PBHs begin to interact with nonlinear structures.

The structure of minihalos shows a cuspy profile with a density running with the radius as $\rho \sim r^{-9/4}$~\cite{Berezinsky:2013fxa, Adamek:2019gns}. Indeed, this internal structure can be readily derived from Eqs.~(\ref{mh}) and (\ref{rh}) as~\footnote{Equation~(\ref{rho}) agrees with Ref.~\cite{Berezinsky:2013fxa} up to a numerical factor of 2.}
\begin{align}
\rho_{\textrm{halo}}(r)& = \frac{1}{4\pi r^2} \frac{dM_{\textrm{halo}}(r)}{dr}\\
&\simeq 6 \times 10^{-21}\,\frac{\textrm{gr}}{{\textrm{cm}}^{3}}\left( \frac{\textrm{pc}}{r} \right)^{9/4} \left( \frac{M_{\textrm{PBH}}}{10^2\,M_{\odot}}  \right)^{3/4}\,,\label{rho}
\end{align}
which is valid for $r_{\textrm{min}} \leq r \leq R_{\textrm{halo}}$, where $r_{\textrm{min}} = 8G_NM_{\textrm{PBH}}$. Such a steep profile was confirmed by N-body numerical simulations performed in~\cite{Adamek:2019gns}.

The ratio of the axion de Broglie wavelength $\lambda_{\text{DB}} = h/(m_a v_a)$ to the halo radius is given by
\begin{equation}
\frac{\lambda_{\text{DB}}}{R_{\text{halo}}} \sim 6 \times 10^{-11} \left( \frac{10^{-5}\,\text{eV}}{m_{\text{a}}} \right) \left( \frac{M_{\odot}}{M_{\text{PBH}}}  \right)^{2/3} \left( \frac{1+z}{1000} \right)^{7/6}\,,
\label{particlelike}
\end{equation}
where we have taken $v_{a} \sim (G_N M_{\text{halo}}/R_{\text{halo}})^{1/2}$ as an estimate of the axion virial velocity. When $\lambda_{\text{DB}}/R_{\text{halo}} \gtrsim 1$  the wave nature of the dark matter particle cannot be ignored, and the accretion should not be efficient. 

\section{Axion Stars Nucleation}

Lattice simulations performed by Levkov \textit{et al.}~\cite{Levkov:2018kau} show axion stars may nucleate kinetically in virialized dark matter halos/axion miniclusters through gravitational BEC.\footnote{In~\cite{Levkov:2018kau}, axions are considered as nonrelativistic bosons which interact themselves via gravitation neglecting the axion self-interaction.} At large occupation numbers, the system is described by a random classical field which evolves under its own gravitational potential. The kinetic regime require to satisfy the following conditions:  
\begin{align}
(m_a v_a) \times (R_{\text{halo}}) \gg 1\,,\label{Cond1}\\
(m_a v_a^2) \times (\tau_{\text{gr}}) \gg 1\,.\label{Cond2}
\end{align}
Here $\tau_{gr}$ is the condensation time scale for the axion star formation. This time scale is proportional to the inverse of the kinetic relaxation rate $\Gamma_{\text{kin}} \sim n_a  \sigma_{\text{gr}}v_a \mathcal{N}$, where $n_a$ is the halo axion number density, $\sigma_{\text{gr}}\approx 8\pi m_a^2 G_N^2 \Lambda /v_a^4$ is the scattering cross section due to gravitational interaction, and $\mathcal{N}=(6\pi^2 n_a) / (m_a v_a)^3$ is the occupancy number related to Bose enhancement. Here $\Lambda \equiv \text{log}_e(m_a v_a R_{\text{halo}})$ is the Coloumb logarithm. 

This relaxation rate differs from the other gravitational rate which appears in classical field theory within the so-called condensation regime, $\Gamma_{\text{cond}} \sim 8 \pi G_N m_a^2 n_a / k^2$ where $k$ is some characteristic wave number~\cite{Guth:2014hsa, Erken:2011dz}. Usually the condensation relaxation rate is larger than the kinetic relaxation rate since $\Gamma_{\text{cond}}$ scales like $G_N$ but $\Gamma_{\text{kin}}$ scales like $G_N^2$.

 In the kinetic regime, $\tau_{gr}$ is calculated to be~\cite{Levkov:2018kau}
\begin{align}
 \tau_{\text{gr}} &  = \frac{b\sqrt{2} m_a^3 v_a^6}{12 \pi^3 G_N^2 \rho_{\text{halo}}^2 \Lambda}\,,\\
\bar{\tau}_{\text{gr}} & \simeq  \frac{4 \sqrt{2}}{27 \pi}  \left(\frac{R_{\text{halo}}}{v_a}\right) (R_{\text{halo}}m_av_a)^3\,,\label{time} 
\end{align}
where $\bar{\tau}_{\text{gr}}\equiv \tau_{\text{gr}} \Lambda$. The  numerical coefficient $b=\mathcal{O}(1)$ depends on the details of the process. To obtain Eq.~(\ref{time}), we have taken $b=1$ and $v_a^2 \sim (4\pi/3)G_N \rho_{\text{halo}}R_{\text{halo}}^2$. 

Even though we will use Eq.~(\ref{time}) as our standard time scale for axion stars nucleation,  this time needs to be considered with caution.  Numerical results reported in~\cite{Eggemeier:2019jsu} show stars nucleation in axion miniclusters occurs at least $\sim\mathcal{O}(10^2)$ times earlier than the time scale predicted by $\tau_{\text{gr}}$. This situation suggests that the true relaxation rate places somewhere between $\Gamma_{\text{cond}}$ and $\Gamma_{\text{kin}}$.  
 
We apply Eqs.~(\ref{Cond1}), (\ref{Cond2}), and (\ref{time}) to analyze axion stars nucleation in dark minihalos of dressed PBHs. 
For numerical calculations, we consider a flat $\Lambda$CDM cosmology and used values based on Planck TT,TE,EE+lowE+lensing+BAO at the 68$\%$ confidence levels in~\cite{Aghanim:2018eyx}. Using  Eqs.~(\ref{mh}), (\ref{rh}) and $v_a \sim (G_N M_{\text{halo}}/R_{\text{halo}})^{1/2}$, we reexpress Eqs.~(\ref{Cond1}), (\ref{Cond2}), and (\ref{time})
as 
\begin{align}
\frac{m_a v_a R_{\text{halo}}}{10^{\alpha}} & \simeq  \left( \frac{m_a}{10^{-5}\,\text{eV}} \right) \left( \frac{1000}{1+z} \right)^{7/6} \left( \frac{M_{\text{PBH}}}{10^{-10}M_{\odot}} \right)^{2/3}\,,\label{mavaR}\\
\frac{m_a v_a^2 \bar{\tau}_{\text{gr}}}{10^{\beta}} & \simeq  \left( \frac{m_a}{10^{-5}\,\text{eV}} \right)^4\left(  \frac{1000}{1+z} \right)^{14/3} \left(   \frac{M_{\text{PBH}}}{10^{-10}\,M_{\odot}}\right)^{8/3} \,,\label{mava2}\\
\frac{\bar{\tau}_{\text{gr}}}{10^{\gamma}} & \simeq  \text{Gy}  \left( \frac{m_a}{10^{-5}\,\text{eV}}  \right)^3  \left( \frac{1000}{1+z}  \right)^5   \left( \frac{M_{\text{PBH}}}{10^{-10}\,M_{\odot}} \right)^2\,.\label{taugr}
\end{align}
where 
\begin{align}
\alpha &\equiv \text{log}_{10}(2.1 \times 10^4)\,,\\
\beta &\equiv \text{log}_{10}(1.3 \times 10^{16})\,,\\
\gamma &\equiv \text{log}_{10}(2.5 \times 10^7)\,.
\end{align}
Define $(A_{\text{min}}, B_{\text{min}}, C_{\text{min}})$ such that the kinetic regime is satisfied at a given redshift, axion and PBH masses according to 
\begin{align}
(m_a v_a R_{\text{halo}}) = 10^{C_{\text{min}}}\,,\label{one}\\
(m_a v_a^2 \bar{\tau}_{\text{gr}})=10^{B_{\text{min}}}\,,\label{two}\\
(\bar{\tau}_{\text{gr}})=10^{A_{\text{min}}}\,\label{three} \text{Gyr},
\end{align}  
Using these expressions in Eqs.~(\ref{mavaR}), (\ref{mava2}), and (\ref{taugr}), we can find the dependence of $(A_{\text{min}}), (B_{\text{min}}), (C_{\text{min}})$ with respect to the variables of interest as follows
\begin{align}
& A_{\text{min}}  = \gamma - \frac{3}{4} \beta + \frac{3}{4}B_{\text{min}} + \text{log}_{10}\left[\frac{10^3}{(1+z)}\right]^{3/2}\,,\label{mat1}\\
& C_{\text{min}} =\alpha - \frac{\beta}{4} + \frac{B_{\text{min}}}{4}\,,\label{mat2}\\
& \text{log}_{10}\left[\frac{\left(\frac{m_a}{10^{-5}\,\text{eV}}\right)^{3/2}}{\left(\frac{10^{-10}\,M_{\odot}}{M_{\text{PBH}}}\right)}\right]  =  -\frac{3}{8}\beta + \frac{3}{8}B_{\text{min}}-\text{log}_{10}\left[ \frac{10^3}{1+z}\right]^{7/4}\,.\label{mat3}
\end{align}
At a given redshift $z$, once you set a value for  $B_{\text{min}}$, Eqs.~(\ref{one}), (\ref{two}), and (\ref{three}) are immediately set together to the linear relation between $M_{\text{PBH}}$ and $m_a$ shown in Eq.~(\ref{mat3}). We see that
as we go deep in the kinetic regime, e.g. as $B_{\text{min}}$ increases, the condensation time increases as well as the related
axion mass ($M_{\text{PBH}}$) for fixed $M_{\text{PBH}}$ ($m_a$). 

\begin{figure}[ht!]
\centering
\includegraphics[scale=0.62]{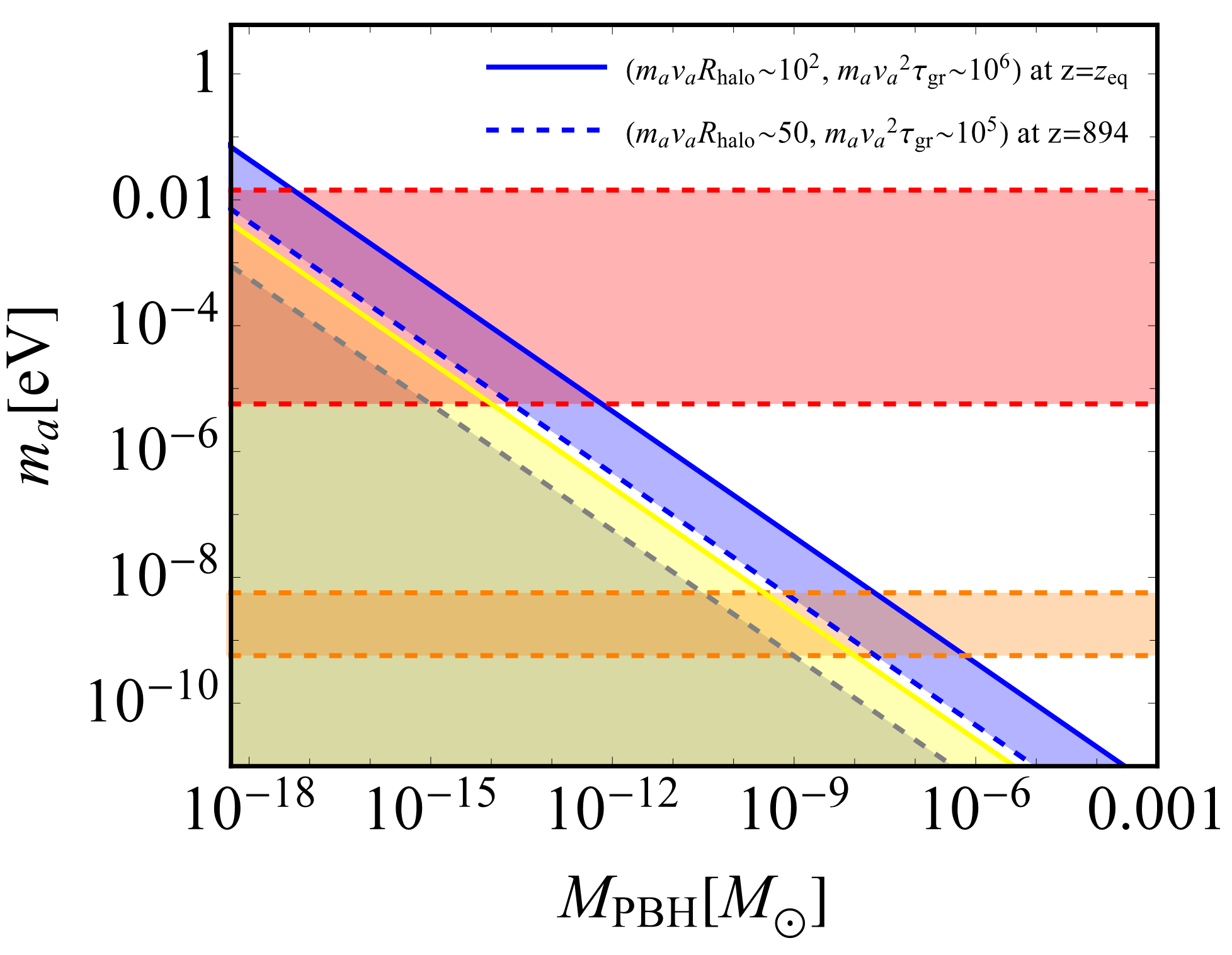}\!\!\!\!\!\!\!\!\!\!\!\!\!
\caption{Contour-levels of $(m_av_aR_{\text{halo}},m_av_a^2\tau_{\text{gr}})$ at a given redshift $z$ in the parameter space ($M_{\text{PBH}},m_a$). The blue shaded region between the blue solid ($z=z_{\text{eq}}$) and dashed  ($z=894$) lines corresponds to the parameter space of ($M_{\text{PBH}},m_a$)  at $894 < z < z_{\text{eq}}$, which satisfies the kinetic regime as $(m_av_aR_{\text{halo}},m_av^2\tau_{\text{gr}}) \gtrsim (50,10^5)$.  The intersection between the red (orange) band and the blue shaded region corresponds to the parameter space for the QCD (string) axion, where $4\times 10^{8}\,\text{GeV} \gtrsim F_a \gtrsim 10^{12}\,\text{GeV}$~\cite{Raffelt:2006cw, PDG:2018} (where $10^{15}\,\text{GeV} \gtrsim F_a \gtrsim 10^{16}\,\text{GeV}$~\cite{Kawasaki:1997ct, Svrcek:2006yi}). The yellow solid (gray dashed) line plus the yellow (gray) shaded region correspond to the zone in which  $\lambda_{\text{DB}}/R_{\text{halo}} \geq 1$ at $z_{\text{eq}}$ (at $z=894$) [see Eq.~(\ref{particlelike})].}
\label{Plot1}
\end{figure}

Using previous equations, we show in Fig.~\ref{Plot1} contour-levels of $(m_av_aR_{\text{halo}},m_av_a^2\tau_{\text{gr}})$ at a given redshift $z$  in  the parameter space $(M_{\text{PBH}},m_a)$.  Between the two conditions required to be within the kinetic regime, Eqs.~(\ref{Cond1}), (\ref{Cond2}), the former condition is the most difficult to achieve. We have taken $(m_av_a)\times(R_{\text{halo}}) \simeq 50$ as the minimum value to consider the system in the kinetic regime. 
 Blue solid ($z=z_{\text{eq}}$) and dashed  ($z=894$) lines correspond to $(m_av_aR_{\text{halo}},m_av_a^2\tau_{\text{gr}})\sim(10^2,10^6)$ and  $\sim(50,10^5)$, respectively. The blue shaded region between these two lines shows the parameter space of ($M_{\text{PBH}},m_a$) at $894 < z < z_{\text{eq}}$ satisfying the kinetic regime as   $(m_av_aR_{\text{halo}},m_av_a^2\tau_{\text{gr}}) \gtrsim (50, 10^5)$.
Even though minihalos at redshift $z \lesssim 894$ can satisfy the kinetic regime for certain parameter space $(M_{\text{PBH}}, m_a)$, we do not include them in Fig.~\ref{Plot1} because their associated condensation time scales lead to nucleations  
after the first galaxies formation, e.g. $z_{\star} \lesssim 30$. At that time, we expect dressed PBHs begin to interact with nonlinear structures so that Eqs.~(\ref{mh}), (\ref{rh}) are no longer valid. Indeed, the whole parameter space
shown in the blue shaded region is associated with condensation time scales ranging as $\tau_{\text{gr}}(z)\sim(10^{-2}-10^{-1})\,\text{Gyr}$, e.g. nucleation in minihalos occurs at redshift $30 \lesssim z_{\star} \lesssim 115$.

The intersection between the red (orange) band and the blue shaded region show the parameter space $(M_{\text{PBH}},m_a)$ associated with axion stars nucleation for the case of the QCD (string) axion. For the QCD axion, PBHs with masses $5 \times 10^{-19}\,M_{\odot}\lesssim M_{\text{PBH}} \lesssim 7\times10^{-13} M_{\odot}$ are able to form an axion dark minihalo which satisfies the kinetic regime at a given redshift $894 \leq z \leq z_{\text{eq}}$. For the case of the string axion, the central PBHs are heavier with a mass range of  
 $7\times10^{-10}M_{\odot} \lesssim M_{\text{PBH}} \lesssim 7\times 10^{-7}\,M_{\odot}$. 

The yellow solid (gray dashed) line plus the yellow (gray) shaded region show the parameter space at $z = z_{\text{eq}}$ ($z=894$) in which the accretion of axion dark matter from PBHs is not efficient, e.g. $\lambda_{\text{DB}}/R_{\text{halo}} \geq 1$ [see Eq.~(\ref{particlelike})]. We see the particlelike behavior of the QCD and string axions in the parameter space of interest is strong enough to ensure their accretion from PBHs. For example, $\lambda_{\text{DB}}/R_{\text{halo}}(z_{\text{eq}}) \sim 10^2$ on the contour-level shown by the blue solid line in Fig.~\ref{Plot1}. Indeed, the axion particlelike behavior is a requirement to satisfy the kinetic regime as shown in Eq.~(\ref{Cond1}).

Numerical simulations in~\cite{Eggemeier:2019jsu} showed axion stars nucleate in local density maxima of axion miniclusters. In particular, they reported two axion stars nucleation in one axion minicluster. As we mentioned earlier, the steep density profile of minihalos runs with the radius as
$\rho \sim r^{-9/4}$~\cite{Berezinsky:2013fxa,Adamek:2019gns}.  Thus, we expect the star nucleation mostly occurs in inner shells of the minihalo at a radius $r_{\text{halo}}$, such that the enclosed minihalo mass is dominant over the PBH mass (e.g., the minihalo gravitational potential is dominant over the PBH potential). The fraction of the minihalo radius which encloses $k$ times the PBH mass can be estimated as $r_{\text{halo}}/R_{\text{halo}} \sim (k M_{\text{PBH}}/M_{\text{halo}})^{4/3}$. Following results in Fig.~\ref{Plot1}, the nucleation of axion stars occurs at $30 \lesssim z_{\star}  \lesssim 115$.  Take $k = 10$. Thus, we expect nucleation in minihalos occurs in spherical shells at a distance from the central PBH of $0.05 R_{\text{halo}}(z_{\star}\simeq 30)\lesssim r_{\text{halo}}\lesssim 0.28 R_{\text{halo}}(z_{\star}\simeq115)$.

In the case that nucleated axions stars correspond to excited states coming from radial perturbations, we expect they tend to settle down at the ground state configuration by ejecting part of the axion particles to eliminate the excess of kinetic energy~\cite{Seidel:1993zk}.\footnote{By using different initial conditions, authors in~\cite{Eggemeier:2019jsu} found axion stars nucleation in highly excited states with nonradial oscillations.}

Suppose that at redshift $z$, the dark minihalo satisfies conditions for axion stars nucleation with a condensation time $\tau_{\text{gr}}(z)$. Thus, the present average parameter density of axion stars in the proposed scenario is given by
\begin{align}
\Omega_{\star,0} =  \left(\frac{N_{\star} M_{\star}}{M_{\text{PBH}}}\right) \xi^{\text{PBH}}_{\text{DM}}\,\Omega_{\text{DM},0}\,,
\label{numberdensity}
\end{align}
where  $N_{\star}$ is the average number of axion stars per halo after nucleation, $\Omega_{\text{DM},0}$ is the present dark matter parameter density, $\xi^{\text{PBH}}_{\text{DM}} \equiv \Omega_{\text{PBH}}/\Omega_{\text{DM}}$ is the fraction of dark matter in PBHs, and $M_{\star}$ is the characteristic mass of axion stars. The present average parameter density shown in Eq.~(\ref{numberdensity}) needs to be taken as an upper bound due to we are not taking into account disruptive events acting on axion stars after their formation.  

Axion stars after nucleation continue capturing axions from the halo until the growth rate slow downs and saturates~\cite{Levkov:2018kau, Eggemeier:2019jsu}. We estimate this mass after saturation by equating the virial velocity of the halo at the nucleation time, e.g. $v_a \sim (G_N M_{\text{halo}}/R_{\text{halo}})^{1/2}$, to the virial velocity in the gravitational potential of the axion star, e.g. $v_{\star} \simeq (G_N M_{\star}m_a/\hbar)$~\cite{Hui:2016ltb}, according to 
\begin{equation}
\left(\frac{M_{\star}}{M_0}\right) \simeq (1+z_{\star})^{1/2} \left(\frac{M_{\text{halo}}}{M_0}  \right)^{1/3}\,,
\label{scaling1}
\end{equation}
where $M_0 \simeq 5.5 \times 10^{-19}\,M_{\odot}\, (10^{-5}\,\text{eV}/m_a)^{3/2}$. 
Up to a numerical factor of order 1, the same scaling relation was previously found for solitonic cores in halos of fuzzy dark matter~\cite{Schive:2014hza}. By using Eq.~(\ref{mh}),  we can rewrite Eq.~(\ref{scaling1}) in terms of the mass of the central PBH to obtain  $(M_{\star}/\overline{M}_0) \simeq (1+z_{\star})^{1/6} (M_{\text{PBH}}/\overline{M}_0)^{1/3}$, where
$\overline{M}_{0} = \sqrt{3000} M_0$. Since this mass shows a very weak dependence on the redshift, we take $M_{\star} \sim (M_{\text{PBH}}^{1/2} \overline{M}_0)^{2/3}$ as the characteristic mass of axion stars in Eq.~(\ref{numberdensity}).

Axion stars nucleation could occur simultaneously in an inner spherical shell of the dark minihalo, so that $N_{\star} \geq 1$ in Eq.~(\ref{numberdensity}). Take, for example, a QCD axion star with $M_{\star} \sim 10^{-18} M_{\odot}$ and $m_a \sim 6\times10^{-4}\,\text{eV}$, being nucleated at $z_{\star} \simeq 115$
within a dressed PBH with $M_{\text{PBH}} \sim 2 \times 10^{-16} M_{\odot}$. Since the axion star radius is given by the relation $R_{\star} \simeq 2 \,\text{km} (10^{-10} M_{\odot}/M_{\star})(10^{-5}\,\text{eV}/m_a)^2$~\cite{Schive:2014hza}, we have $R_{\star}/(0.28 R_{\text{halo}}) \sim 2\times 10^{-3}$ and $M_{\star}/M_{\text{halo}} \sim  2 \times10^{-4}$. There are enough space and mass to consider multiple nucleation in an inner shell.      
Therefore, using Eq.~(\ref{numberdensity}), the present average parameter density of axion stars is estimated as  
\begin{equation}
\Omega_{*,0} \sim N_{\star} \left(  \frac{\overline{M_0}}{M_{\text{PBH}}} \right)^{2/3} \xi^{\text{PBH}}_{\text{DM}}\Omega_{\text{DM},0}\,,\label{rangeomegastar}
\end{equation}
where $N_{\star} \geq 1$.

\begin{figure}[ht!]
\centering
\includegraphics[scale=0.28]{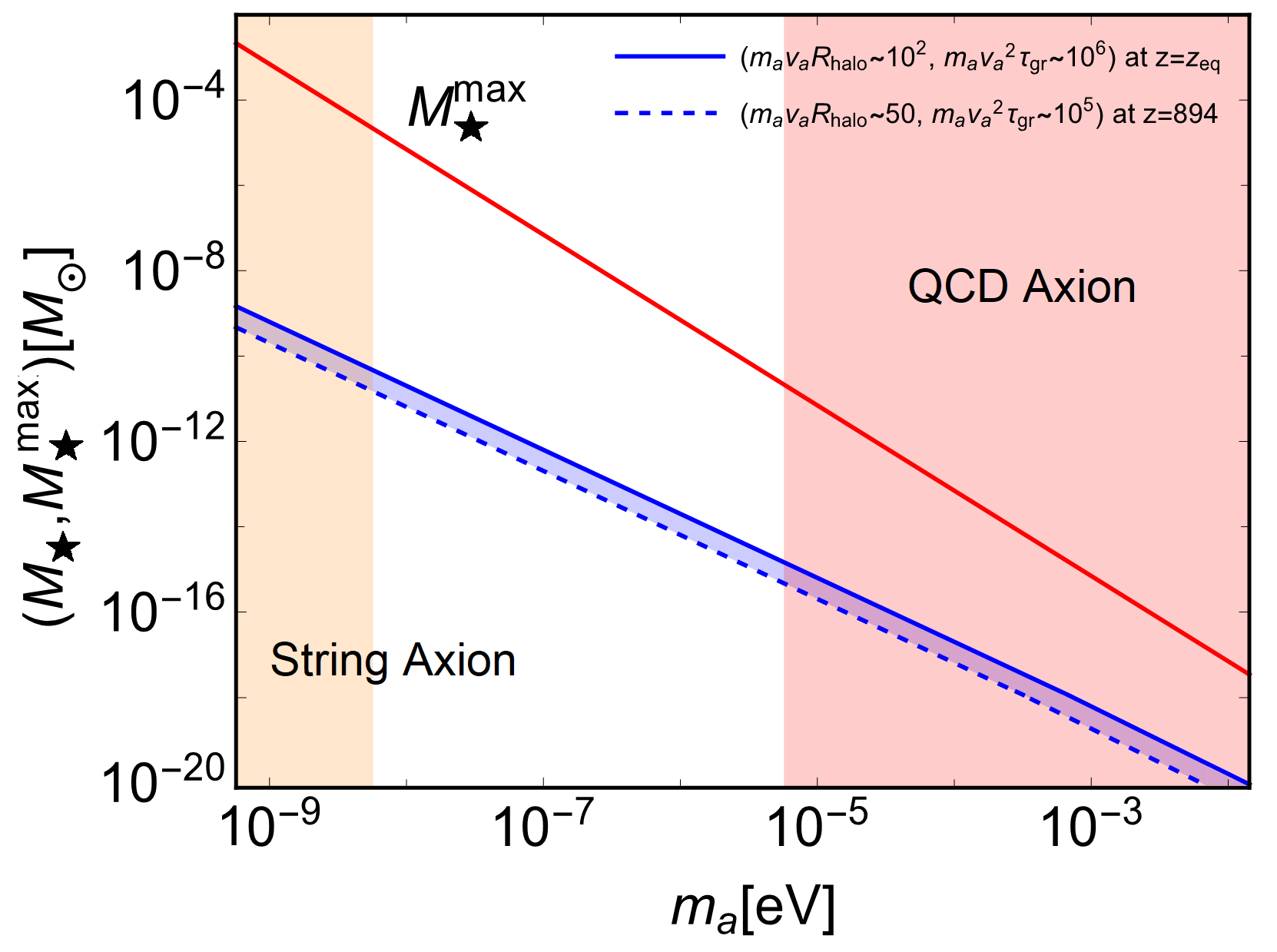}\!\!\!\!\!\!\!\!\!\!\!\!\!
\caption{ Contour-levels of ($m_av_aR_{\text{halo}},m_av_a^2\tau_{\text{gr}}$) in the parameter space $(m_a,M_{\star})$.  The blue shaded region between the blue solid ($z=z_{\text{eq}}$) and dashed  ($z=894$) lines corresponds to the parameter space of $(m_a,M_{\star})$ at which the kinetic regime is satisfied as $(m_av_aR_{\text{halo}},m_av^2\tau_{\text{gr}}) \gtrsim (50,10^5)$ at $894< z < z_{\text{eq}}$. Red (orange) band corresponds to the mass range for the QCD (string) axion. The red solid line indicates the theoretical maximum mass, $M_{\star}^{\text{max}}$, that an axion star in the ground state configuration can achieve~\cite{Schiappacasse:2017ham}.}
\label{Plot3}
\end{figure}

Figure~\ref{Plot3} shows contour-levels of  $(m_av_aR_{\text{halo}},m_av_a^2\tau_{\text{gr}})$ in the parameter space $(m_a,M_{\star})$ for the QCD (red band) and the string  (orange band) axion case. The blue solid $(z=z_{\text{eq}})$  and dashed $(z=894)$ lines refer to the values $(m_av_aR_{\text{halo}},m_av_a^2\tau_{\text{gr}})\sim(10^2, 10^6)$ and $\sim(50, 10^5)$, respectively.  The blue shaded region between these two lines refers to the parameter space of $(m_a,M_{\star})$ at which the kinetic regime is satisfied as $(m_av_aR_{\text{halo}},m_av^2\tau_{\text{gr}}) \gtrsim (50,10^5)$ at $894< z < z_{\text{eq}}$.  For the case of the QCD axion, axion stars masses after saturation range as $\mathcal{O}(10^{-20})\,M_{\odot} \lesssim M_{\star} \lesssim  \mathcal{O}(10^{-15})\,M_{\odot}$. When minihalos are composed by the string axion, axion stars masses are heavier ranging as $\mathcal{O}(10^{-11})\,M_{\odot} \lesssim M_{\star} \lesssim \mathcal{O}(10^{-9})\,M_{\odot}$. In principle, $M_{\star}$ can still grow after the saturation point although at a very suppressed rate~\cite{Eggemeier:2019jsu}. Note that values of $M_{\star}$ are far away to reach (and overpass) the  maximum mass allowed for an stable axion star configuration,  $M^{\text{max}}_{\star} \sim 7 \times 10^{-12}\,M_{\odot} (10^{-5}\,\text{eV}/m_a)^2$~\cite{Schiappacasse:2017ham}. Thus, their collapse and explosion in relativistic axions~\cite{Levkov:2016rkk} is unlikely.

The mass of PBHs can be associated with observational constraints leading to an upper bound on the fraction of dark matter that PBHs can explain, $\xi^{\text{PBH}}_{\text{DM,max}}$~\cite{Carr:2009jm, Barnacka_2012, Graham:2015apa, Niikura:2017zjd, Tisserand:2006zx, Inomata:2017okj}. In particular, PBHs with masses in the range $5\times10^{-19}\,M_{\odot} \lesssim M_{\text{PBH}} \lesssim 5 \times 10^{-17} \,M_{\odot}$ emit a significant amount of photons which contribute to the extragalactic photon background. EGRET and FERMI constrain this mass range in terms of $\xi^{\text{PBH}}_{\text{DM,max}}$~\cite{Carr:2009jm}. The absence of any femtolensing events from gamma-ray bursts of known redshift constrains PBHs with masses $5 \times 10^{-17}\,M_{\odot} \lesssim M_{\text{PBH}} \lesssim 10^{-14}\,M_{\odot}$~\cite{Barnacka_2012}. Existence of white dwarfs in our local galaxy constrains PBHs with masses $5\times 10^{-15}\,\lesssim M_{\text{PBH}} \lesssim 10^{-13}$~\cite{Graham:2015apa}. Null observations of microlensing events using the Subaru HSC data constrain PBHs with masses   $ 10^{-13}\,M_{\odot} \lesssim M_{\text{PBH}} \lesssim 10^{-6}\,M_{\odot}$~\cite{Niikura:2017zjd}.

\begin{figure}[ht!]
\centering
\includegraphics[scale=0.62]{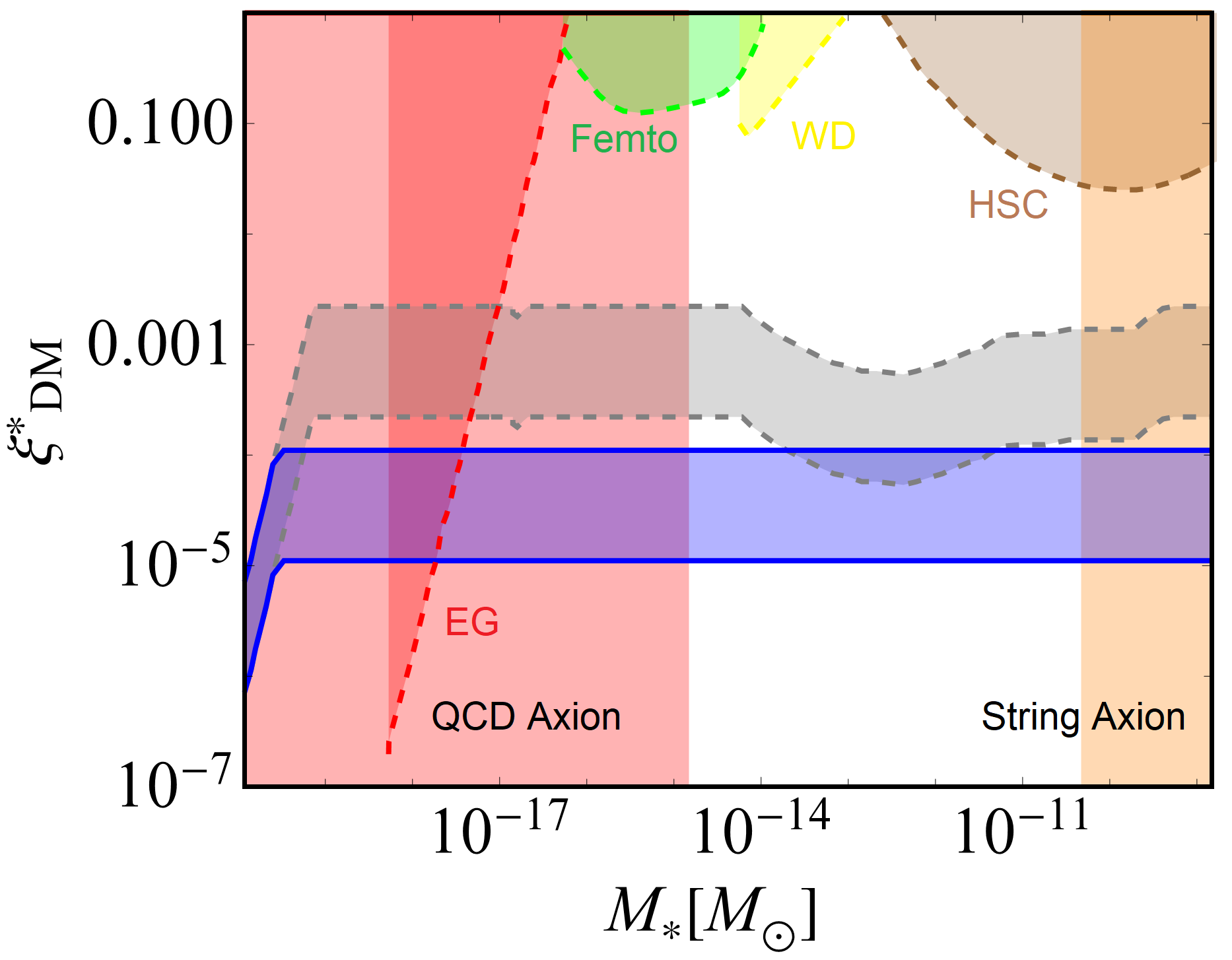}\!\!\!\!\!\!\!\!\!\!\!\!\!
\caption{The blue (gray) shaded band shows the estimate of the current fraction of dark matter in axion stars $\xi^{\star}_{\text{DM}}$ by using Eq.~(\ref{rangeomegastar}), a fraction in PBHs no greater than $0.5\%$ ($10\%$),  $1 \leq N_{\star} \leq 10$, and the contour-level $(m_av_aR_{\text{halo}},m_av_a^2\tau_{\text{gr}}) \sim (10^2,10^6)$ at $z = z_{\text{eq}}$ in the parameter space $(M_{\text{PBH}},M_{\star})$.
Light red (light brown) band corresponds to the mass range of axion stars associated with the QCD (string) axion. In addition, we have shown constraints over the PBH abundance. In particular, extragalactic photon background (EG)~\cite{Carr:2009jm}, femtolensing (Femto)~\cite{Barnacka_2012}, white dwarfs in our local galaxy (WD)~\cite{Graham:2015apa}, and Subaru HSC data (HSC)~\cite{Niikura:2017zjd}.}
\label{Plot2}
\end{figure}

 We consider in Eq.~(\ref{rangeomegastar}) a conservative  maximum fraction of dark matter in PBHs as 
 $\xi_{\text{DM}}^{\text{PBH}} = \text{min}(\xi^{\text{PBH}}_{\text{DM},\text{max}},\text{Q})$, where $\text{Q}$ is predefined. The blue and gray shaded regions in Fig.~\ref{Plot2} show an estimate of the current fraction of dark matter in axion stars, $\xi^{\star}_{\text{DM}}$. As we mentioned before, here we are not considering tidal disruptive events acting on axion stars after nucleation. We have used the contour level of  $(m_av_aR_{\text{halo}},m_av_a^2\tau_{\text{gr}})\sim (10^2,10^6)$ at $z=z_{\text{eq}}$,  and $1 \leq N_{\star} \leq 10$.
The blue and gray shaded regions consider a  fraction of dark matter in PBHs no greater than $0.5\%$ and $10\%$, respectively (e.g., $\text{Q} = 5\times10^{-3}\, \text{and}\, 10^{-1}$, respectively).
Light red and light brown bands correspond to axion star masses associated with the QCD axion and the string axion case, respectively.  
We have showed in addition constraints over PBHs coming from extragalactic photon background (EG), femtolensing (Femto), white dwarfs (WD), and Subaru HSC microlensing (HSC). Note that constraints over PBHs abundance lead to constraints over nucleated axion stars, specially for $\text{Q}$ sufficiently large as shown the irregular shape of the gray shaded region.

Numerical simulations are required to determine with accuracy the relation $M_{\star} = M_{\star}(M_{\text{halo}})$, the parameter space $(M_{\text{PBH}},m_a)$ associated with axion stars nucleation, and the average number of  nucleated axion stars per halo.

\section{Discussion}

Suppose conditions in dark minihalos for formation of axion stars are satisfied such that they are formed before the time of first galaxies formation, $z_{\star} \sim 30$.  After nucleation of axion stars in minihalos, the fate of axion stars and the central PBH is a rather complicated scenario. While some of axion stars may leave minihalos, others may remain within them. An eventual merger between axion stars and the central PBH is also an interesting possibility which deserves future analysis.

Since nucleation would happen in the inner shells of minihalos, tidal forces coming from the mean field potential of the dressed PBH may disrupt these compact objects after their formation. While the tidal radius $(r_{\text{tidal}})$ is much less than the radial distance of the axion star from the central PBH, we may safely apply the distant-tide approximation. The tidal radius is calculated to be~\cite{2008gady.book.....B}
\begin{equation}
r_{\textrm{tidal}} = \left(  \frac{M_{\star}}{3 M_{\textrm{PBH}_d}(R_0)} \right)^{1/3} [f_{\text{PBH}_d}(R_0)]^{-1/3} R_0\,,\label{rtid}
\end{equation} 
with
\begin{equation}
f_{\text{PBH}_d}(R_0) = 1 - \frac{1}{3} \frac{d\text{ln}M_{\text{PBH}_d}(r)}{d\text{ln}r} \Bigr|_{R_0}\,, 
\end{equation}
where $R_{\textrm{0}}$ is the radial distance of the nucleated axion star from the central PBH and  $M_{\text{PBH}_d}(R_0)$ is the total mass of the dressed PBH within a radius $R_0$. The mass profile of the dressed PBH is estimated as  
\begin{equation}
M_{\text{PBH}_d}(r) = 4\pi \int_{r_{\text{min}}}^{r}dr'\,r'^2\, \rho_{\text{halo}}(r') + M_{\text{PBH}}.
\end{equation}
 Let us consider again a typical QCD axion star
with $M_{\star}\sim 10^{-18}\,M_{\odot}$ and $R_{\star} \sim 2 \times 10^{-9}\,\text{pc}$, which  has been nucleated at a redshift $z_{\star}\simeq 115$ within a dressed PBH with $M_{\text{PBH}} \sim 2\times 10^{-16} M_{\odot}$, $M_{\text{halo}} \sim 6\times 10^{-15}\,M_{\odot}$ and $R_{\text{halo}} \simeq 3 \times 10^{-6}\,\text{pc}$. As we explained in the previous section, we expect nucleation of the axion star occurs in inner shells of the minihalo, such that $R_0 \sim 0.28 R_{\text{halo}}$ where the minihalo potential is dominant over the PBH potential. Using Eq.~(\ref{rtid}), we have  $r_{\text{tidal}}/R_{\star} \simeq 30$ at $R_0  \sim 0.28 R_{\text{halo}}$, so that the axion star is resistant against tidal disruption from the mean field potential of the dressed PBH.

The likelihood of a merger between the nucleated axion star and the PBH may be estimated by calculating the free fall time 
between both astrophysical objects as $t_{\text{ff}} \sim (\pi/2) R_0^{3/2}/\sqrt{2 G_N (M_{\text{PBH}})}$, where we have used $M_{\text{PBH}} \gg M_{\star}$. For the case studied above, we have $t_{\text{ff}} \sim \,\text{Myr}$, which is a long time. 
Tidal forces acting over the dressed PBH coming from, for example, encounters with other dressed PBHs, may lead to the expulsion of the nucleated axion star from the minihalo.

In any case, we expect galactic halos at the time of formation around $z \sim 6$ would be composed by isolated and clustered dressed PBHs (containing axion stars), naked PBHs, and axion stars as well as
smooth axion dark matter background. Axion stars within galactic halos will undergo different level of disruption mainly coming from the mean field potential of the galaxy, disk shocking, and  encounters with stars.

We discuss in~\cite{Hertzberg:2019exb} the possibility of a significant part of the dark matter background ends up localized in the form of minihalos around PBHs within galactic halos and effects of this on dark matter direct detection (see~\cite{ losHeros:2020csi} for the general status of dark matter searches).
The nucleation of axion stars in these dark minihalos addresses in this article would complement this picture.

Obtaining a more complete understanding of the proposed scenario, specially about the fate of axion stars after nucleation, requires to perform a set of numerical simulations, which is beyond the scope of the present article. We leave this task for future work.

The local dark matter density around a few hundred parsecs around the Sun is $\rho_{\text{DM}}^{\text{local}} \sim 0.3\, \text{GeV\,cm}^{-3}$. Thus, the total local number of axion stars can be expressed as
\begin{equation}
N^{\text{total}}_{\star, \text{local}} \simeq 10^{11}\,\left(\frac{\xi^{*}_{\text{DM}}}{0.03}\right) \left(\frac{10^{-11}\,M_{\odot}}{M_{\star}}\right) \left(\frac{r}{100\,\text{pc}}\right)^3\,.\label{Ntotal}
\end{equation}
From Fig.~\ref{Plot3}, take $M_{\star} \sim 10^{-18}\,M_{\odot}$ ($M_{\star} \sim 10^{-10}\,M_{\odot}$) as the typical mass of axion stars for the QCD (string) axion. Taking a conservative $0.5\%$ in the fraction of dark matter in axion stars, we have up to $\sim10^{17}$ ($10^{9}$) axion stars around our Sun in a radius of $100\, \text{pc}$ for the QCD (string) axion.\footnote{This estimate needs to be taken with cautious since we are not considering tidal disruption events in the Milky Way.}

The number of encounters between the Earth and an axion star per unit of time is calculated as
\begin{equation}
N_{\bigotimes\star} = n^{\text{local}}_{\star,0} \sigma_{\text{eff}} v_{\text{rel}}\,,
\end{equation}
where $n^{\text{local}}_{\star,0} = \xi^{\star}_{\text{DM}} \rho_{\text{DM,local}}/ M_{\star}$ is the local number density of axion stars, $\sigma_{\text{eff}}$ is the geometrical cross section for the encounter between the Earth and an axion star enhanced by gravitational focusing, and $v_{\text{rel}} \simeq 3 \times 10^2 \,\text{km/s}$ is the relative velocity between both astrophysical objects. 

Taking a conservative $0.5\%$ in the fraction of dark matter in axion stars with a typical mass  $M_{\star} \sim 10^{-18}M_{\odot}\, (M_{\star}\sim 10^{-10}M_{\odot})$ for the QCD (string) axion case, the number of encounters results to be 
$N_{\bigotimes \star} \sim 10^{-1}\,\text{Myr}^{-1}\, (N_{\bigotimes \star} \sim 10^{-3}\,\text{Myr}^{-1})$. Hence chances of direct detection of dark matter by the Earth passing through an axion star is extremely small.

However, if a non-negligible number of axion stars survive tidal disruptions, then their presence today within the Milky Way halo would enhance DM indirect detection experiments.\footnote{See~\cite{Iwazaki:2014wka, Raby:2016deh, Dietrich:2018jov} for astrophysical signatures coming from collisions between axion and neutron stars and~\cite{Hertzberg:2018zte} for photon emission via parametric resonance.}  

\section{Acknowledgments}
 T. T. Y. is supported in part by the China Grant for Talent Scientiﬁc
Start-Up Project and the JSPS Grant-in-Aid for Scientiﬁc Research Grants No.
16H02176, No. 17H02878, and No. 19H05810 and by World Premier
International Research Center Initiative (WPI Initiative), MEXT, Japan.
M. P. H. is supported in part 
by National Science Foundation Grant No. PHY-1720332.\\
$^*$mark.hertzberg@tufts.edu\\
$^\dagger$enrico.e.schiappacasse@jyu.fi\\
$\ddagger$tsutomu.tyanagida@ipmu.jp

\bibliographystyle{apsrev4-1} 
\bibliography{AxionStars_12072020} 

\end{document}